\documentclass[journal]{IEEEtran}

\usepackage{cite}
\usepackage{comment}
\usepackage{bm}
\usepackage{graphicx}
\graphicspath{{./Figures/}}
\DeclareGraphicsExtensions{.pdf,.jpeg,.png}
\usepackage{amsmath}
\usepackage{amssymb}
\usepackage{algorithm}
\usepackage{algorithmic}
\usepackage{xcolor}
\usepackage{url}
\usepackage{array}
\usepackage{fancyhdr}
\usepackage{subcaption}
\usepackage{algorithm}
\usepackage{algorithmic}
\usepackage{bm}

\newcolumntype{M}[1]{>{\centering\arraybackslash}m{#1}}

\begin{document}

\title{
Real-Time Cascade Mitigation in Power Systems Using Influence Graph Improved by Reinforcement Learning
}
\author{Kai~Zhou, Youbiao He, Chong Zhong, Yifu Wu

\thanks{K. Zhou is with the School of Mechanical and Electrical Engineering, Soochow University, Suzhou, China (email: kzhou@suda.edu.cn).
Youbiao He was with the Department of Computer Science, Iowa State University. (email: yh54@iastate.edu).
Chong Zhong was with the Department of Mechanical Engineering, University of Akron. (email: cz29@zips.uakron.edu).
Yifu Wu was with the Department of Computer and Information Technology, Purdue University. (email: wu1584@purdue.edu)
}
}
\maketitle

\begin{abstract}
Despite high reliability, modern power systems with growing renewable penetration face an increasing risk of cascading outages. Real-time cascade mitigation requires fast, complex operational decisions under uncertainty.
In this work, we extend the influence graph into a Markov decision process model (MDP) for real-time mitigation of cascading outages in power transmission systems, accounting for uncertainties in generation, load, and initial contingencies. The MDP includes a do-nothing action to allow for conservative decision-making and is solved using reinforcement learning.  
We present a policy gradient learning algorithm initialized with a policy corresponding to the unmitigated case and designed to handle invalid actions. The proposed learning method converges faster than the conventional algorithm. Through careful reward design, we learn a policy that takes conservative actions without deteriorating system conditions. 
The model is validated on the IEEE 14-bus and IEEE 118-bus systems. The results show that proactive line disconnections can effectively reduce cascading risk, and certain lines consistently emerge as critical in mitigating cascade propagation. 
\end{abstract}
\IEEEpeerreviewmaketitle

\section{Introduction}
Cascading in electric power transmission systems is a sequence of dependent outages that successively weaken the power system\cite{baldickPES08}. It is triggered by initial events and then propagates. Some cascades, especially the longer ones, will result in a blackout with significant amounts of load shedding, whereas others do not result in load shedding and can be regarded as precursors to a blackout. These blackouts are rare but high-impact events that occur often enough to pose a substantial risk to society \cite{hinesEP09,carrerasPS16}. 
Cascading outages involve various mechanisms and are highly stochastic phenomena, especially as renewable energy resources increasingly penetrate modern power systems. It remains a great challenge to mitigate cascade risks, especially in real-time. 

Effective mitigation strategies for operational or emergency response to extreme events require fast actions within limited decision-making windows. With the advancement of smart grid technologies, modern power systems are equipped with advanced automation and control capabilities, making fast operational response feasible. 
On the other hand, the study of the real outage data in \cite{dobsonPS21} reveals that, in cascading events, outages occur almost simultaneously and are categorized into generations; the typical interval between these generations is 10 to 20 minutes. This temporal structure offers a limited but critical window during which rapid decision-making and mitigation strategies can be deployed to contain the propagation of cascading outages.

Due to the inherent complexity and stochastic nature of cascading outages, the Markov property is widely adopted in related studies. Influence graph-based methods, also known as interaction graphs, have been proposed to capture the characteristics of cascading propagation and to provide insight for mitigation strategies \cite{hinesPS17, qiPS15, zhouPS20, MahshidEnergies20}. The influence graph model constructs a graph whose nodes are failed system components and whose edges are influence qualifications between component outages. Cascading characteristics are then extracted by analyzing the influence matrix using different methods. The Markov assumption is not explicitly stated in \cite{hinesPS17,qiPS15}, while \cite{zhouPMAPS20} clearly defines a rigorous Markov chain to represent the propagation of transmission line outages driven by real outage data.  
Moreover, a PageRank-based approach, which is effectively a Markov Chain, is employed to screen vulnerable transmission lines in cascading propagation \cite{MaSG19}.
In addition to discrete-time Markov chains, continuous-time Markov processes have also been applied to model the dynamic evolution of cascading outages \cite{Rahnamay-NaeiniPS14, Rahnamay-NaeiniSG16, wangMahshidPS18}. 

Most studies on cascading mitigation focus on the planning horizon by identifying critical components involved in cascades. It is useful to distinguish critical components in initial outages and those during propagation, as the underlying mechanisms are different and mitigation strategies should be applied accordingly. Influence graph-based models have been widely used to identify critical components in propagation by various analytical methods. For example, \cite{zhouPS20} defines a Markovian influence graph and estimates the probability that transmission lines are involved in large cascades. Similarly, \cite{sunPS24} defines a stochastic influence graph and applies eigen-analysis to identify critical components based on high participation factors. Other methods aim to break key linkages that have high weights in the interaction graph to localize the outage propagation \cite{chenPS23,qiPS15,hinesPS17}.      
However, these mitigation strategies are not variant for different events or initial contingencies, and are implemented in advance rather than during cascading events in real time. 


A key challenge of real-time cascading risk mitigation lies in the scale and complexity of system states, as well as uncertainties associated with extreme events and initial outages. 
\cite{sunAccess19} proposes an optimal power flow method for on-line cascading mitigation. DC optimal power flow is augmented with constraints that involve key vulnerable components to cascading.
However, conventional optimization methods based on detailed system models are often computationally intensive and thus may be impractical for emergency control situations. To enable timely decision-making, approximations are needed to reduce the computation burden. Moreover, mitigation strategies are limited to a certain set of strategies; otherwise, broader strategy spaces are hard to optimize in real time. 

An alternative approach involves training a model offline that can be deployed for real-time use. 
Deep reinforcement learning (DRL) demonstrates promising results for emergency control scenarios \cite{huangSG20,huangPS22}. It is a sequential decision-making process and is capable of exploring large action spaces. 
Several studies have explored the application of reinforcement learning in cascading mitigation. 
\cite{zhangPS20} proposes a method based on reinforcement learning to search for representative risky fault chains, which are frequent cascading segments in cascading outages. A reinforcement learning framework is constructed to search for and identify these fault chains offline, which are then utilized online through knowledge transfer. 
In addition to offline learning, RL methods have also been developed for online decision-making during cascade propagation. \cite{dasSJ22} presents a Markov decision process (MDP) framework to determine load-shedding actions to mitigate cascading outages. However, the power system state is simplified using abstracted states, defined by the number of failed lines, the maximum capacity among them, and a binary variable indicating whether the cascade ends. 
\cite{zhangRESS24} introduce a DRL-based method that enables operators to take remedial actions during the propagation of cascading outages. Remedial actions are disconnecting line intentionally to cut off the propagation path. While the model accounts for uncertainties in initial contingencies, it assumes static generation and load conditions, which may limit its practical applicability. 

In this paper, we extend the influence graph model by relaxing the assumption that the influence only exists between outaged components in successive generations (where a generation of a cascade is a set of simultaneous outages). Instead, the influences are from all previous outages, considering a broader and more realistic range of cascading interactions. We also include mitigation actions and formulate the problem as a Markov decision process. A do-nothing action is explicitly introduced as a null strategy, which serves as a way of making conservative intervention. Moreover, we consider the variability in load and generation, which fluctuates continuously. It is impractical to train separate models for all possible system states, particularly when these involve continuous variables. We aim to generalize across a wide range of operating conditions. 
The contributions of this paper are summarized as follows:
\begin{itemize}
    \item Formulate a Markov decision process model for cascading risk mitigation under uncertainty and introduce a do-nothing action to enable cautious decision-making;
    \item Solve the MDP model using deep reinforcement learning with a pretrained policy and consideration of invalid actions;
    \item The study shows that the intentional disconnection of transmission lines can mitigate the risk of cascading, and that critical lines vary by system state but typically constitute a small subset of all transmission lines. 
\end{itemize}


\section{The MDP formulation for cascading mitigation}
As introduced in \cite{zhouPS20}, a cascade $\bm X$ is modeled as a discrete Markov chain, denoted as $X_0, X_1, ...,X_t,..., X_{T}$, where $X_t$ is the set of simultaneous line outages in generation $t$ and $T$ is the total number of generations. 

We relax this assumption of state representation by including all previous line outages and the percentages of line flow in transmission lines. For distinguishing, we denote the state as $S_t$ instead of $X_t$. 
\begin{align}
    S_t = [l_1,...,l_n, \rho_1,...,\rho_n]
\end{align}
where $n$ is the number of transmission lines, $l_i$ is a binary variable indicating whether line $i$ is connected ($l_i = 1$) or not, $\rho_i = P_{l_i} / P_{l_i,max}$ is the percentage of active power flow $P_{l_i}$ of line $i$. 
Other features can also be added to $S_t$, such as generation and loads. The termination state $S_T$ is a collection of states with no new outages, indicating the end of cascades. 

The time difference between consecutive generations $S_t$ and $S_{t+1}$ in historical observed outages ranges from 10 to 20 minutes \cite{dobsonPS21}. This time window can be used by a fast decision-making method to offer mitigation strategies. Let $A_t$ denote the strategy or action taken after state $S_t$, then the cascade with a mitigation strategy is represented as $S_0, A_0, S_1, A_1, ...,S_t, A_t..., S_{T}$. Then, it is natural to model the cascade with a Markov decision process, denoted as 
\begin{align}
   \bm X = S_0, A_0, S_1, A_1, ...,S_t, A_t..., S_{T}.
\end{align}

The transition between states is determined by the power flow solution and described as
\begin{align}
    p(S_{t} | S_{t-1}, A_{t-1}), \qquad t = 1,...,T. \label{equ:dyn}
\end{align}
If state $S_{t-1}$ includes all variables that the power flow solver needs, such as the node power injection, bus voltage magnitudes and angles, and line status, then the transition (\ref{equ:dyn}) degenerates into a deterministic transition. However, since the power system state is not fully observable, (\ref{equ:dyn}) is stochastic. 

Action $A_t$ may include active load shedding, generation dispatching, islanding by disconnecting tie lines, reconnecting available lines, etc. Load and generation are continuous variables, while line statuses are binary variables. Due to the complexity of different types of actions, we only intentionally consider line disconnection as actions at our disposal in this paper. Also, it is impossible to disconnect any number of lines once in practice. For simplicity and without losing generality, we assume that only one line can be disconnected in each generation. 

We emphasize the necessity of introducing a do-nothing action. First, to mitigate the cascade, we should be conservative in making any operation unless we are highly positive that the operation does reduce cascading risk. Second, some actions may have negligible or no effect on system states; in such cases, we should do nothing to avoid unnecessary operation cost. Finally, an agent may take an action that attempts to disconnect an already outaged line, and a do-nothing action should replace this action. 

The action at any generation is given by a probability distribution conditioned on state $S$, which is called a policy 
\begin{align}
   \pi(A=a|S=s)(a \in \{0,...,n\}, \forall s) 
\end{align}
where $A=0$ represents the do-nothing action and $A=i$ represents disconnecting line $l_i$. The concept of an agent refers to an entity that learns an optimal policy and takes actions accordingly. 

We aim to optimize the policy $\pi(A|S)$ that minimizes cascading risk. Cascading risk is quantified based on the number of generations, the number of line outages, and the relative amount of load shedding. The negative immediate impact at each step serves as the reward in the MDP. Thus, maximizing the total reward in the MDP is equivalent to minimizing the overall cascading risk.

Generation 0 corresponds to the system state with initial outages. Initial outages are caused by external factors, such as bad weather, which cannot be mitigated during cascading propagation. Hence, the immediate impact of generation 0 is not considered in the reward. Given the initial system state, the MDP agent will determine an action to mitigate the risk of possible cascading, This action could be a do-nothing action, which is the optimal option when the cascade stops. If the cascade propagates to generation 1, the reward for this action is calculated as the sum of -1, the impact in terms of new line outages and load shedding, and any applicable action penalty, as shown in (\ref{equ:reward}). The number of new line outages is transformed with an exponential function so that the typical value is less than 1. Thus, the impacts of generation number, line outage number, and load shedding have close weights such that no one is dominating the reward.  
If the cascade continues to propagate, the evaluation of the reward for other generations is the same. Moreover, if the power flow fails to converge, we apply a large penalty $-100$ to avoid this situation. Usually, there is a cost for operating the power system, so we penalize for taking proactive actions by adding a small negative number $-\alpha$ to the reward. It also serves as a way of favoring the do-nothing action over other actions when they result in the same system state. In addition, when the action is disconnecting an outaged line, it is replaced by a do-nothing action and penalized.
$\alpha$ should be less than $100 \left(1 - e^{-0.01} \right)$ so that the reward of proactively disconnecting a line is greater than a line outage. 

Therefore, the reward at generation $t$ is: 
\begin{align}
    R_t(S_{t-1}, A_{t-1}, S_t) = & - I_{S_t \neq S_T} -100 I_{fail} - \alpha I_{A_{t-1} \neq 0} \notag \\
    & - 100 \left(1 - e^{-0.01 N_{g,t}} \right)  \notag \\
    & - (L_{t-1} - L_t)/L_{t-1}  \label{equ:reward}
\end{align}
where $I$ is an indicator function, $\alpha \in (0,0.99)$, $N_{g,t}$ is the number of new line outages after taking action in generation $t$, $L_t$ is the total amount of load at step $t$. $I_{S_t \neq S_T} $ evaluates $1$ when $S_t$ is not the termination state and $0$ otherwise, and $I_{fail}$ is $1$ when the power flow does not solve. The impact of line outages increases monotonically with the range $[0,100)$. The Taylor series of this impact shows that it is close to $N_{g,t}$ but strictly less than $N_{g,t}$. This makes the model prefer to stop the cascading early when the total number of line outages is the same. For example, if a cascade has 3 line outages in generation 1 and then it stops, then the return (total reward of this cascade) is about $-4$ ($-1$ for propagating to next generation, $-3$ for line outages in generation 1); while if an action (e.g. disconnecting one line) is taken and results in 1 line outage in generation 1 and 1 line outage in generation 2, then the return is $-5$ (the reward for generation 1 is $R_1 = -3$ due to propagation, the line outages and taking action, and the reward for generation 2 is $R_2=-2$ due to propagation and line outages). 

Return, which is the total reward of a cascade, is denoted as $G(\bm{X}) = \sum_{t=1}^T R_t$. It is the negative of cascading risk. Then, the goal of the MDP problem is $\max_{\pi} E_{\pi}[G(\bm{X})]$.
where $E_{\pi}$ is the expectation given that the action is taken following policy $\pi$. 
\looseness=-1

In summary, a cascade is modeled as an MDP. States particularly focus on transmission lines, actions correspond to operations made at each generation, such as proactive transmission line disconnection, and rewards are the negative impact of line outages in each generation. Transitions between states are determined by the power flow solution, which is stochastic because system states are not fully observable. 
The decision-making flow is: 
\begin{enumerate}
    \item Initialize the power system with initial outages.
    \item Take an action. \label{action}
    \item Evaluate the system state.
    \item Calculate the corresponding reward.
    \item If cascade is not stopping, propagate outages, go to \ref{action}); otherwise, terminate the process.    
\end{enumerate}

\section{Cascading simulation with mitigation actions}
\label{sec:casmodel}
\begin{figure}[!ht]
    \centering
    \includegraphics[width=\columnwidth]{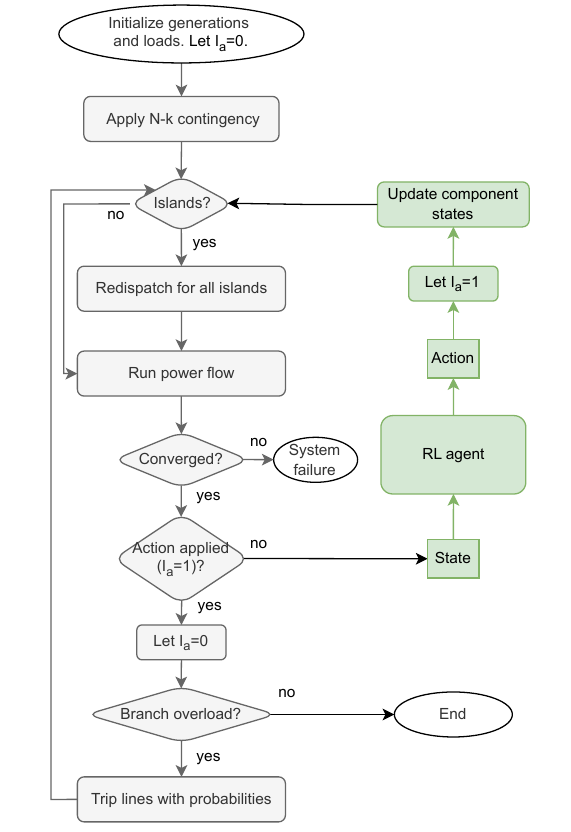}
    \caption{The simulation flowchart of power system cascading with mitigation actions.}
    \label{fig:cascademodel}
\end{figure}

There are various power system cascading models to simulate the cascade propagation, such as the OPA model with fast and slow timescales \cite{dobsonchaos07}. For our application, we only need the fast timescale process to simulate the cascading propagation; however, we need to add the function of applying mitigation actions. All of these cascade models make assumptions about the propagation mechanism to some extent. This paper uses a slightly modified version of the cascading model described in \cite{hinesPS12}. This cascading simulation model can easily be replaced with other models to better simulate the power system dynamics or particular propagation mechanisms. The flowchart of simulating a cascade with actions is shown in Figure \ref{fig:cascademodel}, and we briefly describe it here:
\begin{enumerate}
    \item Initialize the simulation by reading system states from generation and load curves, and sample initial outages according to the distribution specified by the contingency list. 
    \item Determine whether an island has formed. If not, proceed to step \ref{powerflow}; otherwise, proceed to the next step. \label{island}
    \item Balance generations and loads in each island: if an island contains no generators or loads, remove all buses in this island from the power flow calculation; otherwise, make generators ramp up or down to reach power balance as closely as possible; then, if generation exceeds load, reduce each generator's output proportionally to its capacity; otherwise, if load exceeds generation, shed load proportionally.
    \item Run power flow. \label{powerflow}
    \item If the power flow does not converge, record a system failure and end this cascade; otherwise, go to the next step.
    \item The agent selects an action based on the policy and updates the power system state. The cascade model then handles any resulting islands and re-runs the power flow analysis. 
    \item If no lines are overloaded, stop the cascade; otherwise, go to the next step.
    \item For each overloaded line, trip it with a probability $\beta_i$, which depends on the relative line flow. If $\rho_i \geq 1.5$, $\beta_i = 1$; if $1 \leq \rho_i < 1.5$, $\beta_i = 2(\rho_i-1)$; if $\rho_i < 1$, $\beta_i = 0$.
    \item Go to step \ref{island}.
\end{enumerate}
 
\section{Illustration with a 5-bus system}
We use the IEEE 5-bus test case to illustrate cascading mitigation based on the MDP model. Since it is a small system, we solve the MDP problem using dynamic programming (DP). 

\begin{figure}[!ht]
    \centering
    \includegraphics[width=0.6\columnwidth]{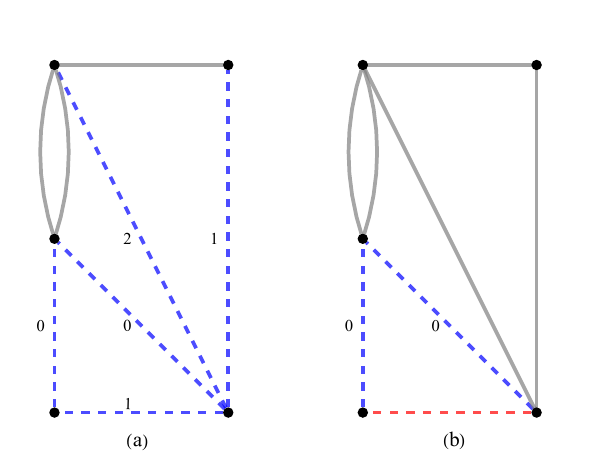}
    \caption{A cascade triggered by outages on line 1 and line 4 (dashed lines labeled with generation 0). Dashed lines are outages and the number along each line indicates the cascade generation. The red line is the proactive disconnection action. (a) Without mitigation. (b) With mitigation.}
    \label{fig:5busresult}
\end{figure}

The topology of the IEEE 5-bus system is shown in Figure~\ref{fig:5busresult}. The system has two generators, three loads, and eight transmission lines; the parameters can be found in \cite{pandapowerPS18}. 
For illustrative purposes and to make the problem tractable for the dynamic programming approach, we make the following assumptions: (1) the state only contains line status $l$ so that it is finite, (2) generations and loads are fixed so that we can enumerate all possible states and obtain the transition dynamics $p(S_t|S_{t-1}, A_t)$, (3) the initial outaged lines are line 1 and line 4.  

\begin{figure}[!ht]
    \centering
    \includegraphics[width=0.5\columnwidth]{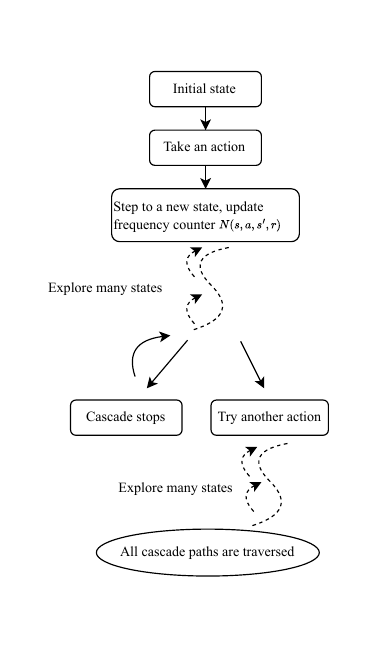}
    \caption{DFS with backtracking used to derive MDP dynamics.}
    \label{fig:dfs}
\end{figure}

We propose the following depth-first search (DFS) method to enumerate all possible states to derive the MDP dynamics $p(S_t|S_{t-1}, A_{t-1})$. The basic idea, as shown in Figure~\ref{fig:dfs}, is: (1) start with initial outages; (2) take the first action and move to the resulting state; (3) from this new state, continue selecting an action and move to the next state until reaching the stop state, and record the frequency of each $(s,a,s',r)$ tuple; (4) backtrack to the previous state and explore any of the remaining actions; (5) repeat until all states and actions have been visited; (6) normalize the frequency of $(s')$ given $(s,a)$ to obtain $p(S_t|S_{t-1}, A_{t-1})$.

After we have $p(S_t|S_{t-1}, A_{t-1})$, we use the DP-based policy iteration algorithm to obtain the optimal policy $\pi_*$. This algorithm first evaluates the value function of the current policy according to the Bellman equation. The value function, denoted as $V_{\pi}(s)$, is defined as the expected return given initial state $s$ under a policy $\pi$. Then, it improves this policy by greedily updating actions with respect to the current value function according to the policy improvement theorem, and repeats this process until the policy is stable. We refer to chapter 4 of \cite{suttonBook18} for details.  

The optimal policy is given by $\pi_*(A = 1 | S = s_0) = 1$, $\pi_*(A = 3 | S = s_1~or~s_2~or~s_3) = 1$, and $\pi_*(A = 0 | S = s_r) = 1$, where $s_0 = [0,1,0,0,1,0,0,0,0]$, $s_1 = [0,1,0,1,1,0,0,0,0]$, $s_2 = [0,1,0,1,1,1,0,0,0]$, $s_3 = [0,1,0,1,1,0,1,0,0]$ and $s_r$ are all other states. This policy will prevent the line 1 and line 4 outages from propagating to other places. The result is shown in Figure \ref{fig:5busresult}~(b). This optimal policy successfully stops cascading with the cost of disconnection of one line. 

In this example, the policy is expressed as a conditional probability distribution. For a large system with numerous states or continuous states, it would be challenging. Hence, a practical option is to approximate the policy by neural networks that map from states to the probability of actions. 

\section{Solving MDP by reinforcement learning}
Given that the MDP dynamics $p(S_t|S_{t-1}, A_{t-1})$ is usually unknown, the state and action spaces could be huge, and the state may involve continuous variables, a practical solution to the MDP problem is the deep reinforcement learning method by learning from cascade data. Therefore, a deep reinforcement learning method is applied in our cascading mitigation MDP problem. Specifically, we solve the MDP problem based on one of the policy gradient algorithms, Proximal Policy Optimization (PPO) \cite{schulmanArxiv17}. Instead of directly applying the algorithm to our problem, we have made some adjustments to accelerate the convergence speed. Figure \ref{fig:algorithm1} shows the flowchart of the proposed learning method. 

The probability of a trajectory (that is a cascade in our application) is: 
\begin{align}
    p_{\bm{\theta}}(\bm{X}) & = p(S_0) p(A_0 | S_0) p(S_1 | S_0, A_0)... \notag \\
    & = p(S_0) \prod\nolimits_{t=0}^{T} \pi_{\bm \theta}(A_t|S_t) p(S_{t+1}|S_t, A_t)
\end{align}
where $p_{\bm{\theta}}(\bm{X})$ is the probability of cascade $\bm{X}$ under a policy network with parameter $\bm{\theta}$. A policy network, denoted as $\pi_{\bm \theta}$, is modeled as a multilayer perception neural network:
\begin{align}
    h^{(l)} & = \sigma(W^{(l)} h^{(l-1)} + b^{(l)}), \quad l=1,...,L \\
    \pi_{\bm \theta} & = \text{softmax}(h^{(L)}) \label{equ:policy}
\end{align}
where $h^{(0)} = s$, $L$ is the number of layers, $\sigma$ is the ReLU activation function, $\bm \theta$ represents all parameters $W^{l}$ and $b^{l}$.

We have defined a reward for each step, and the objective is to maximize the expected total reward of a trajectory. 
\begin{align}
     \mathbb{E}(G(\bm{X})) = \overline{G} = \Sigma_{n=1}^{N} G(\bm{x}^{(n)  }) p_{\bm{\theta}}(\bm{x}^{(n)}) 
\end{align}

To maximize the total reward, we can use the gradient ascent. It turns out that 
\begin{align}
    \nabla \overline{G} = \frac{1}{N} \Sigma_{n=1}^{N} G \left( \bm{x}^{(n)} \right) \Sigma_{t=1}^{T_n}\nabla \log{\pi_{\bm{\theta}} \left( a_t^{(n)} | s_t^{(n)} \right)}
\end{align}
where $N$ is the number of cascades, $T_n$ is the length of cascade~$n$. 
In this formulation, we can avoid computing the trajectory probability. It corresponds to maximizing the likelihood of $\pi(\mathcal{\theta})$ weighted by the return $G \left( \bm{x}^{(n)} \right)$ of a trajectory.  

We use PPO algorithm to optimize the expected returns with respect to parameter $\bm{\theta}$. PPO replaces the weight $G \left( \bm{x}^{(n)} \right)$ with a defined advantage estimate $\hat{A_t}$ based on the current value function $V_{\bm{\phi}}(s)$, and uses techniques such as importance sampling and PPO-clip for better learning \cite{schulmanArxiv17}. The objective of the learning algorithm is 
\begin{align}
    \max_{\bm \theta} \frac{1}{N} \Sigma_{n=1}^{N} \frac{1}{T_n} \Sigma_{t=1}^{T_n} \min & \left( \frac{\pi_{\bm \theta} (a_t | s_t)}{\pi_{\bm \theta_{old}} (a_t | s_t)} A_{\pi_{\bm \theta_{old}}}(s_t, a_t), \right. \notag \\
    &\left. \quad g(\epsilon, A_{\pi_{\bm \theta_{old}}}(s_t, a_t))\right)
\end{align}
where $g(\epsilon, A) = (1+\epsilon)A$ when $A\geq0$, $g(\epsilon, A) = (1-\epsilon)A$ when $A<0$, and $\epsilon$ is the clip hyperparameter. The value function $V_{\bm{\phi}}(s)$ is also approximated by a neural network and is referred to as a value network. 

We adapt the standard PPO algorithm in two ways to accelerate learning: (1) by pre-training a policy that always produces the do-nothing action, and (2) by excluding invalid actions during training.

\subsection{Pre-train a do-nothing policy}
PPO initializes the policy network with random parameters. However, cascades without mitigation can be treated as using a policy that always selects the do-nothing action. Since the policy gradient method updates its policy network parameters towards the direction of increasing expected return, it is meaningless to start the improvement of the policy network with a random policy with random parameters. Instead, it is better to initialize the PPO with a policy that only selects the do-nothing action. Thus, the PPO algorithm will learn a policy better than the do-nothing policy that mitigates cascades. 

\begin{figure*}[!tp]
    \centering
    \includegraphics[width=0.7\textwidth]{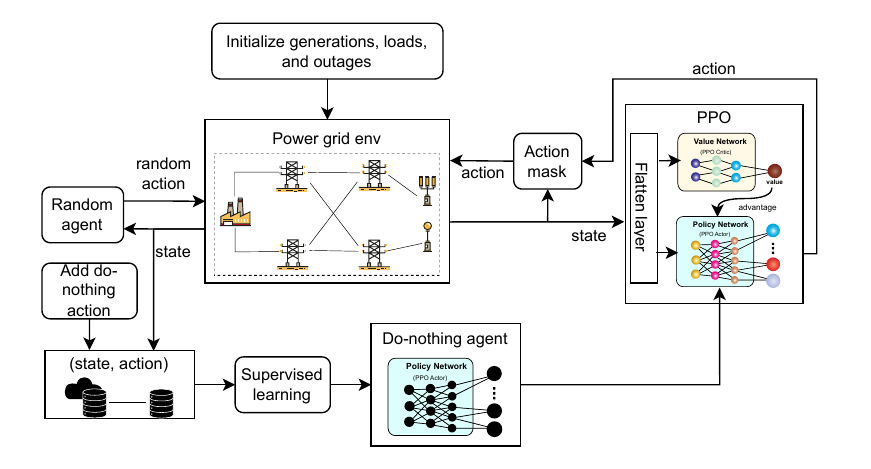}
    \caption{The reinforcement learning framework with a pretrained policy and masking of invalid actions}
    \label{fig:algorithm1}
\end{figure*}

We formulate the learning of a do-nothing policy as a supervised learning problem. The policy network has the same architecture as the one in RL training. Its input is the state we extract from the power environment and its output is the action. The purpose is to learn such a policy that will predict the do-nothing action in any state. The pre-training contains two steps: data collection and training. In the data collection step, we use a random policy, which takes the state as input and outputs a random action, to interact with the power environment and collect states. Note that we do not use a policy predicting the do-nothing action in data collection because the goal of interacting with environment is to explore as many states as possible. As shown in Figure~\ref{fig:algorithm1}, after obtaining a state from the environment, we will pair it with the do-nothing action and save it to the dataset. Then the policy is trained by the supervised learning with the categorical cross-entropy loss, where states are the input features and do-nothing actions are their labels. We include an entropy regularization term in the training objective so that the learned do-nothing policy assigns a relatively high probability to the do-nothing action while maintaining non-negligible probabilities for other actions. This balance is crucial because the later reinforcement learning stage relies on adequate exploration to optimize the policy.

\begin{figure*}[!tp]
    \centering
    \includegraphics[width=0.7\textwidth]{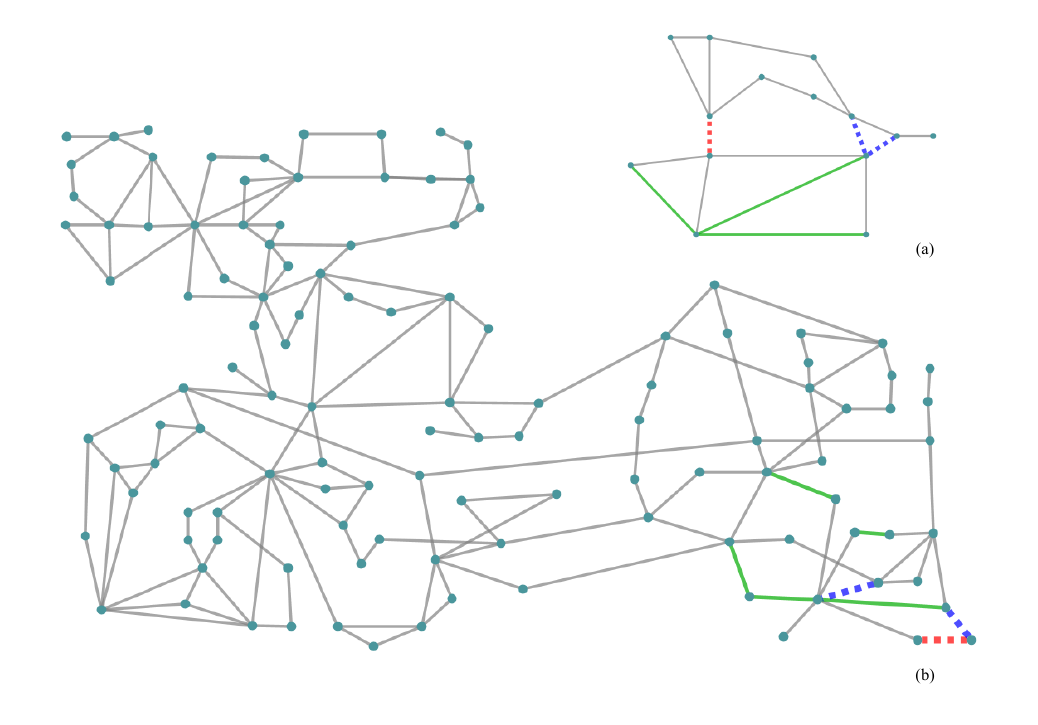}
    \caption{Topologies of the test cases and detailed cascades. Blue dashed lines are outages, red dashed lines are disconnected by the agent, and green lines will fail if no agent takes preventive actions. (a) IEEE 14-bus system. (c) IEEE 118-bus system. }
    \label{fig:cases}
\end{figure*}

\subsection{Masking invalid actions}
If a transmission line has already been disconnected due to an outage or prior operation, it cannot be disconnected again. If the agent tries to operate a disconnected line, this action will be replaced by a do-nothing action. However, this replacement can degrade learning efficiency by exploring many invalid actions and will confuse legitimate do-nothing actions with invalid disconnection attempts. Therefore, we exclude invalid actions and just sample the valid actions. Specifically, the logits corresponding to invalid actions in the policy network are replaced with a large negative number $M$. Let $h_i'^{(L)}$ be the $i$-th logit corresponding to action $i$:
\looseness = -1
\begin{align}
    h_i'^{(L)} =  
    \begin{cases}
        M, & \text{if } l_i = 0 \\
        h_i^{(L)}, & \text{if } l_i = 1 
    \end{cases}
\end{align}

Then the policy (\ref{equ:policy}) is updated to $\pi'_{\bm \theta}  = \text{softmax}(h'^{(L)})$.
Thus, the gradients corresponding to invalid actions become zero; this still produces a valid policy gradient\cite{huangArxiv22}. 

\subsection{Uncertainty consideration}
We consider two sources of uncertainties to generate scenarios with which the agent interacts: loads and generations, as well as initial outages. Loads and generations vary at any time; initial outages are caused by external factors such as lightning or storms.  

A straightforward way to capture the uncertainties from loads and generations is to use a group of time-series scenarios \cite{moralesAP10}. At each time when the environment is reset, we take a snapshot of the current loads and generations, then propagate possible outages based on the cascading model. Various methods can be used to generate scenarios, such as Gaussian mixture models\cite{yangSE2024} and Generative Adversarial Networks\cite{chenPS18}. In this paper, we use publicly available load and generation datasets. 

To better evaluate the risk of cascading, we sample all possible line outages $N-k$ instead of only those that lead to cascades. In practice, when outages occur, operators cannot predict whether they are going to propagate or not; however, they still need to respond to these outages to mitigate potential cascading based on current system states. To reflect this uncertainty in our analysis, we generate initial outages according to contingency motifs. The study of the historical outage data reveals that initial outages occur more frequently in contingency motifs \cite{zhouPS23}. A contingency motif is a subgraph of the system network that occurs significantly frequently. For example, star patterns composed of two, three, or four lines with a common node are contingency motifs that have a higher probability of failure than other types of subgraphs. Hence, we form a contingency list that includes contingency motifs with lines up to $k$ and other random $N-k$. 

\subsection{Summary of the learning process}
To summarize, the learning process begins by initializing the environment with loads and generations at time step $t$, as well as initial outages from the contingency list; then, it trains a do-nothing policy as the initialization of the policy network in the PPO algorithm; finally, it employs the PPO algorithm to train the optimal policy, during which invalid actions are masked. Figure \ref{fig:algorithm1} shows the overall learning framework. 

\section{Case studies}
We test the proposed MDP model and the learning algorithm in IEEE 14-bus and 118-bus test cases, respectively. We use an Ubuntu machine that has an Intel(R) Xeon(R) w5-3425 CPU (3.2 GHz and 12 cores), NVIDIA GeForce RTX 4080 GPU, and 64 GB of RAM. The learning algorithm is implemented in Python using the stable-baseline3 package \cite{raffinStablebaseline21}, the power system cascading environment is constructed on top of the grid2op package \cite{grid2op} and is extended with the proposed failure propagation mechanism described in Section \ref{sec:casmodel}.

\subsection{IEEE 14-bus system}
The IEEE 14-bus system has 20 transmission lines, 6 generators, and 11 loads. The system topology is shown in Figure \ref{fig:cases}(a), and the parameters can be found in \cite{pandapowerPS18}. 

\begin{figure}[!ht]
    \centering
    \includegraphics[width=0.8\columnwidth]{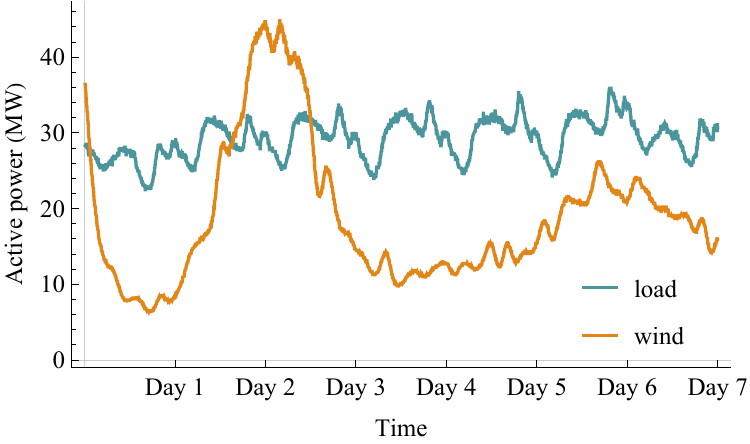}
    \caption{Generation and load curves at two buses. }
    \label{fig:14busgenload}
\end{figure}

To account for the load and generation uncertainty, we use load and generation curves for one week with a resolution of 5 minutes to initialize the states of the power system. A wind power plant and a load at two buses are demonstrated in Figure \ref{fig:14busgenload}.
Since the 14-bus system is small, initial outages are sampled from all $N-1$ as well as $N-2$ with common buses. 

\begin{figure}[!ht]
    \centering
    \includegraphics[width=\columnwidth]{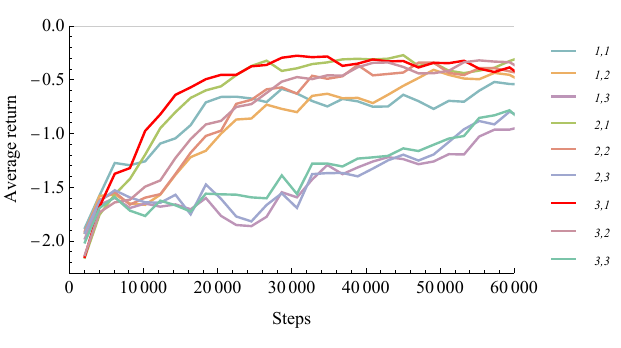}
    \caption{Learning curves for the 14-bus case with different hyperparameters, including entropy coefficients ($[0.1, 0.01, 0.001]$) and learning rates ($[10^{-3}, 10^{-4}, 10^{-5}]$). Legend $i,j$ indicates using $i$-th entropy coefficient and $j$-th learning rate. Returns are smoothed by an exponential moving average to highlight the trends.}
    \label{fig:convergence14}
\end{figure}

We use the grid search method to optimize main hyperparameters such as learning rate and entropy loss coefficient. Three levels of learning rates ($[10^{-3}, 10^{-4}, 10^{-5}]$) and entropy coefficients ($[0.1, 0.01, 0.001]$) are obtained, resulting in a total of nine combinations. Figure \ref{fig:convergence14} shows the convergence curves for different hyperparameters. The selected values are given in Table \ref{tbl:hyperpar}.

\begin{table}[!ht]
\centering
\caption{Hyperparameters of the training algorithm}
\label{tbl:hyperpar}
\begin{tabular}{lcc}
 Hyperparameter & 14-bus case & 118-bus case \\ \hline
 Total number of steps & 60,000 & 600,000 \\
 Learning rate & $10^{-3}$ & $10^{-4}$ \\
 Discount factor & 1 & 1 \\
 Entropy loss coefficient & 0.001 & 0.01 \\
 Clipping of the policy net & 0.2 & 0.3\\
 Clipping of the value net & 0.2 & 0.3\\
 Number of step per update & 1024 & 1024 \\
 Hidden layers of policy net & [64, 64] & [256, 256] \\
 Hidden layers of value net &  [64, 8] & [64, 8] \\ \hline 
\end{tabular}
\end{table}

The trained agent can effectively mitigate the propagation of outages. As shown in Figure~\ref{fig:cases}(a), initial outages are blue, dashed lines. If no action is taken, the green lines would outage in the following generations; however, if the red line is proactively disconnected immediately after initial outages, the cascade will stop here. 

\begin{figure}[!ht]
    \centering
    \includegraphics[width=0.9\columnwidth]{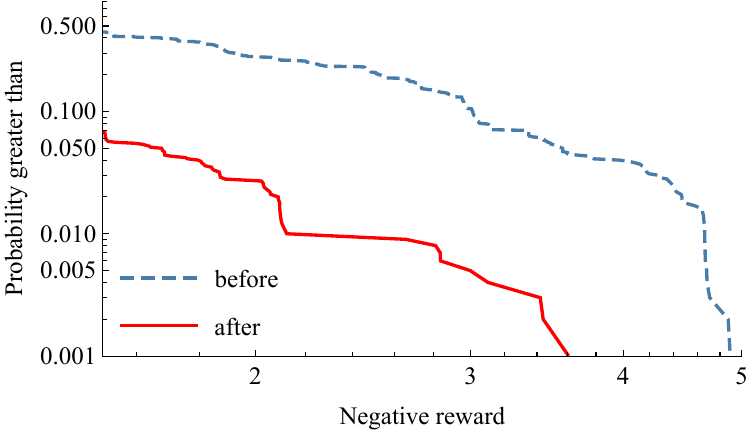}
    \caption{The survival function for negative returns using 1000 cascades with random initial states for IEEE 14-bus system before and after mitigation.}
    \label{fig:14busreward}
\end{figure}

We further test the performance of the trained agent by running $1000$ cascades with randomly selected initial outages, loads and initial outages. Figure \ref{fig:14busreward} shows a comparison of the negative returns before and after the agent's interference. The negative total cascade reward is actually the cascade impact measured by the number of generations, the number of line outages, and relative load shedding. 

\subsection{IEEE 118-bus system}
We further validate the scalability of the proposed method on a larger system, the IEEE 118-bus test case. The network is shown in Figure \ref{fig:cases}(b), and the parameters are from \cite{pandapowerPS18}. 

Similarly to the 14-bus system case study, we reset the environment based on load and generation curves for one week with a resolution of 5 minutes to initialize the states of the power system. 
The initial outages are selected from $N-k$ ($k = 2,3,4$) contingency motifs. 

\begin{figure}[!ht]
    \centering
    \includegraphics[width=0.9\columnwidth]{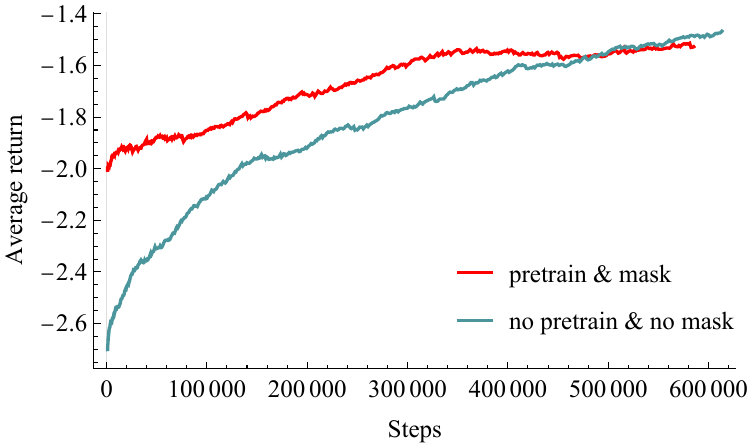}
    \caption{Learning curves for the 118-bus case using PPO-based algorithm with and without pre-training and invalid action masking. Returns are smoothed by an exponential moving average to highlight the trends.}
    \label{fig:convergence118}
\end{figure}

The hyperparameters of the learning algorithm and the RL agent are shown in Table~\ref{tbl:hyperpar}. We compare the proposed learning method with the conventional PPO algorithm. As shown in Figure~\ref{fig:convergence118}, learning with a pre-trained model and consideration of invalid actions converges faster than the conventional PPO algorithm. 

We demonstrate a detailed cascade as shown in Figure~\ref{fig:cases}(b), the cascade triggered by blue lines is successfully stopped by intentionally disconnecting the red line.  

\begin{figure}[!ht]
    \centering
    \includegraphics[width=\columnwidth]{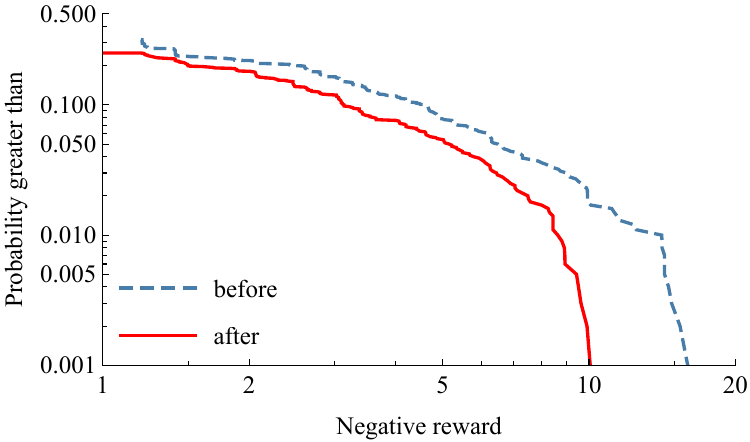}
    \caption{The survival function for total negative rewards using 1000 cascades with random initial states for IEEE 118-bus system before and after mitigation.}
    \label{fig:118busreward}
\end{figure}

The negative returns with and without an agent are shown in Figure \ref{fig:118busreward}, which is successfully reduced by the agent taking proactive actions. 

\begin{figure*}[!tp]
     \centering
     \begin{subfigure}[b]{0.66\columnwidth}
         \centering
         \includegraphics[width=\textwidth]{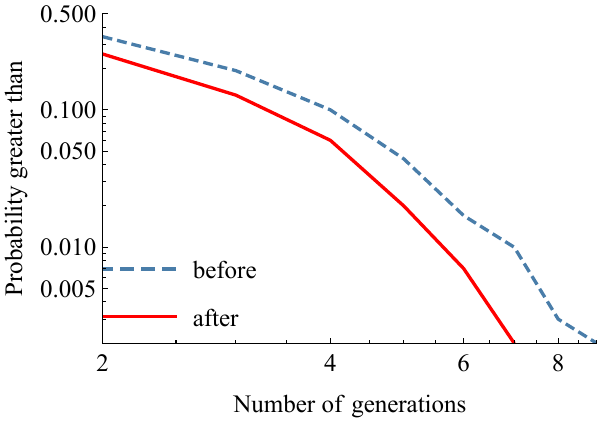}
         \caption{Generations}
         \label{fig:118busmetrics_a}
     \end{subfigure}
     \begin{subfigure}[b]{0.66\columnwidth}
         \centering
         \includegraphics[width=\textwidth]{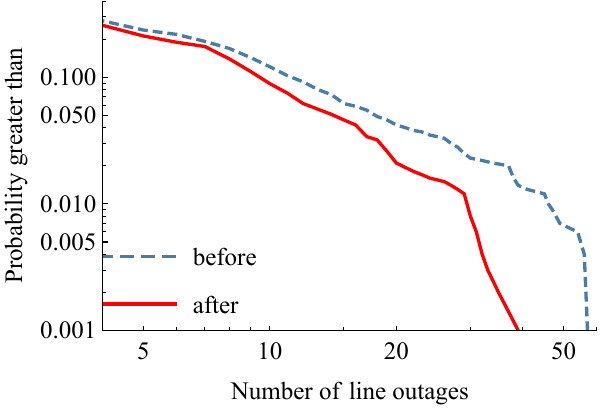}
         \caption{Line outages}
         \label{fig:118busmetrics_b}
     \end{subfigure}
     \begin{subfigure}[b]{0.66\columnwidth}
         \centering
         \includegraphics[width=\textwidth]{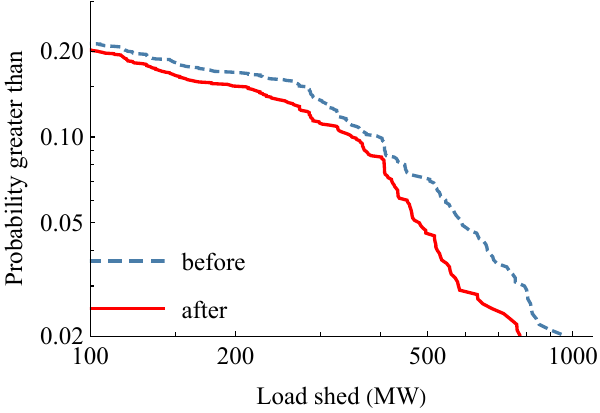}
         \caption{Load shedding}
         \label{fig:118busmetrics_c}
     \end{subfigure}
      \caption{Survival functions for three cascade size metrics using 1000 cascades with random initial states for IEEE 118-bus system before and after mitigation.}
        \label{fig:118busmetrics}
\end{figure*}

We can also assess the mitigation effect on the number of generations, the number of line outages, and the load shedding, respectively, as shown in Figure \ref{fig:118busmetrics}. It shows that the number of generations and line outages is obviously reduced, while the load shedding is slightly reduced. This is because the reward is designed to mitigate the cascading propagation; while load shedding contribution to the reward is less than the others.

\begin{figure}[!ht]
    \centering
    \includegraphics[width=\columnwidth]{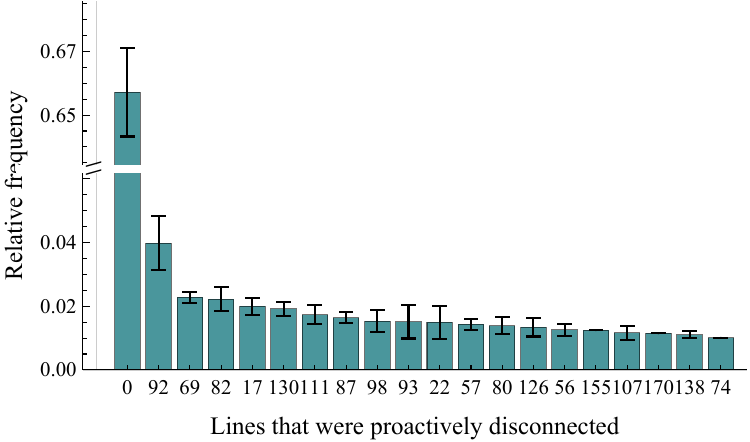}
    \caption{The relative frequency of actions and their standard deviations. The error bars are standard deviations estimated from samples.}
    \label{fig:14busactfre}
\end{figure}

Further, the actions taken by the agent are recorded. Figure \ref{fig:14busactfre} shows the relative frequency of actions and its corresponding standard deviation as error bars. The standard deviation is estimated by testing the agent on multiple batches of initial states.
The most frequent action is do-nothing action. For the rest actions, not all actions share a similar frequency. An interesting finding is that the number of actions with high frequencies ($>0.01$) keeps constant. 

\section{Discussion}
The proposed MDP model represents the system state using line states and relative line flows. Compared to the Markov assumption in influence graph, this extends the state which only contains line outages in a generation. The action space consists of proactive single-line disconnections.  
Despite these simplifications, we demonstrate that the cascade risk can be mitigated by the agent through deep reinforcement learning. A better agent may be possible by including more relevant features of the system states, by taking more actions like generation dispatching, or by training with other algorithms.

Operating the power system stressed by outages should be performed with great care. We prefer an agent that takes action only when it is positive that the cascade can be mitigated. Relative line flow is an important indicator of how stressed the system is; hence, this feature is introduced in the states of MDP formulation. We introduce the do-nothing action and initialize training from a do-nothing policy. When no line is overflow, the agent should do nothing. In practice, a fast detailed simulation can be performed after the agent takes proactive actions to confirm that the actions are valid and do not cause an unexpected result. 
The do-nothing action also enables cascade simulation without mitigation applied. 

The power system cascade model used in DRL environment can be replaced by any other cascade simulation model. It is inherent that any model must simplify the power system dynamics and cascading mechanisms to some extent. As more details are added to the environment, the efficiency of computation becomes a challenge for training. The DRL relies on heavy interactions between the agent and the environment to explore the observation space and action space, which is fundamental to the success of the DRL problem. Interactions may consume a lot of computation time. GPU is not that helpful for speeding up the interactions because it is more suitable in learning neural network parameters in the policy gradient part after we have the data. 

It is not easy to predict whether outages develop into a large cascade or not. Therefore, we sample initial outages from all possible multiple line outages $N-k$ ($1<k\leq K$). To accurately evaluate the cascade risk, we select initial outages in contingency motifs that have high outage probabilities.   

Among the agent's actions, certain lines consistently emerge more frequently than others, suggesting their critical rules in mitigating cascades. This insight allows the mitigation strategy to focus on a small critical subset of transmission lines. Hence, operators can prioritize monitoring and deploy automated devices on these critical lines, which makes it cost-effective and practical.     

\section{Conclusion}
In this paper, we formulate the real-time mitigation of cascading outages in power transmission systems as a Markov decision process. The cascading risk is incorporated into the reward of MDP, which is measured in terms of the number of generations, the number of line outages, and the relative amount of load shedding. 
The action space includes proactive transmission line disconnections as primary control strategies. A do-nothing option is also introduced to allow for conservative decisions when intervention is not justified. The agent is penalized when it tries to disconnect an already outaged line or when the disconnection has the same effect as a do-nothing action. 
This design ensures that the agent disconnects a line only when it can reduce the cascade risk; otherwise, it will do nothing. The do-nothing action also serves the purpose of replacing the agent's action that disconnects an already outaged line.

We present a PPO-based learning algorithm that is initialized with a pre-trained do-nothing policy and masks invalid actions during training. This accelerates the convergence by reducing unnecessary exploration of invalid actions. It is significant as DRL requires extensive agent-environment interactions to explore large observation and action spaces, which is time-consuming. 
The environment accounts for the variability of generation and loads. Initial outages are sampled from contingency motifs to better simulate cascading risk. 

We validate the proposed model on the IEEE 14-bus and IEEE 118-bus systems. The result shows that proactive single-line disconnection can reduce cascading risk and does not deteriorate power system states. The lines disconnected by the agent are not uniformly distributed over all lines; instead, certain lines that appear more frequently than others are critical to prevent failure propagation. The agent determines the disconnection of specific lines in a cascade. 

The proposed model can be a valuable tool for transmission system operators by recommending preemptive actions to mitigate cascading outages before they escalate. 
To improve its practicality, the power system model can be refined and calibrated using historical outage data to ensure alignment with real-world system behavior. Moreover, the action space can be expanded beyond single-line disconnection to include load shedding, generation redispatching, and topology reconfiguration. In practical implementation, it is beneficial to integrate rule-based analysis or DC power flow with the proposed DRL model, as it enables high-stake actions undergo a fast validation process to prevent AI-driven decisions that could introduce unintended risks to the power system. 

\bibliographystyle{IEEEtran}
\bibliography{IEEEabrv,references}

\end{document}